\documentclass[a4paper,10pt]{article}
\usepackage[utf8x]{inputenc}
\usepackage{graphicx}
\usepackage{epsfig}
\title{Thermodynamics and Spectroscopy of Charged Dilaton Black holes}
\author{R Tharanath and  
        V C Kuriakose \\ 
\\
\textit{Department of Physics}\\
\textit{ Cochin University of Science and Technology}\\
\textit{Cochin- 682022,India}}

\begin{document}

\maketitle

\begin{abstract}
The Bohr-Sommerfeld quantization rule is useful to study the area
spectrum of black holes by employing adiabatic invariants. This
method is extended to charged dilaton black holes in 2+1 dimensions.
We put the background space-time into the Kruskal-like coordinate to
find the period with respect to Euclidian time. Also assuming that
the adiabatic invariant obeys Bohr-Sommerfeld quantization rule,
detailed study of area and entropy spectrum has been done. It is
dependent on the charge and is equally spaced as well.
 We also investigate the thermodynamics of the charged dilaton black hole.
\end{abstract}

\section{Introduction}
\label{intro}
One of the most known solutions of Einstein equation is black hole solution. Black holes are the most surprising objects in general relativity.
One of the important characteristics of a black hole is its thermodynamic behaviour. The discovery that the black hole
laws are thermodynamic in nature leads to the notion that there should be an underlying statistical description of them in terms of
microscopic states. Black hole thermodynamics is widely studied \cite{Bekenstein,Hod}. The parameter 'entropy'  which connects thermodynamics and
statistical mechanics scales like the area of the event horizon, unlike its usual role in  thermodynamic systems.
This special behaviour of entropy makes the black hole thermodynamics curious.
Apart from the area law of entropy, the surface area of event horizon shows a quantized behaviour which in turn quantizes the entropy.
The quantization of the black hole horizon area is one of the most fascinating subjects of quantum gravity
ever since Bekenstein first proposed in 1974 that the black hole area spectrum is equally spaced\cite{Bekenstein}. Bekenstein conjectured the possibility of a
connection between quasinormal modes of black holes, which have been studied extensively, and the area spectrum.
By considering minimum change of the horizon area in the process of assimilation of a test particle falling into a black hole,
it was shown that the area spectrum should be linearly quantized, i.e. $A=\gamma m \hbar$, where $\gamma$ is an
undetermined dimensionless constant\cite{Bekenstein} and $m $ is an integer.
An important step in this direction was made by Hod\cite{Hod} in the semi-classical analysis of macroscopic oscillation modes of
black holes.

 Assuming that the horizon area of a black hole behaves as a classical adiabatic invariant, Bekenstein\cite{Bekenstein} later showed
that the quantized area spectrum has the following form
$(\Delta A)_{min}= 8 \pi l_{p}^{2}$ ,where $l_{p}= (\frac{G h}{c^{3}} )^{1/2}$ is the Planck length. Subsequently many attempts have been made to derive the area spectrum and entropy
spectrum directly utilizing the dynamical modes of this classical theory.
Hod found that if one employs the correspondence principle of Bohr, the quantized area spectrum can be determined by
the real part of quasinormal frequencies of the black hole. Hod suggested that the area spectrum is
$(\Delta A)_{min}= 4 ln 3l_{p}^{2}$ .

               Based on the Bekenstein proposal for the adiabaticity of the black hole horizon area and the proposal suggested by Hod
regarding the quasinormal frequencies, Kunstatter\cite{Kunstatter} derived the area spectrum of d-dimensional spherically symmetric black
holes. The specific result for the horizon area quantum is same as obtained by Hod  and by Bekenstein and Mukhanov\cite{mukhanov}.
The work done by Hod is an important step in this direction.
The recent speculation of Maggiore\cite{Maggiore} that the periodicity of a black hole may be the origin of the area quantization law is confirmed.
Zeng\cite{zeng} et al exclusively utilize the period of
motion of an outgoing wave, which is shown to be related to the vibrational frequency of the perturbed black hole, to
quantize the horizon areas of a Schwarzschild and a Kerr black holes. It is shown that the equally spaced area spectrum for both cases takes the same form and the
spacing is the same as that obtained through the quasinormal mode frequencies.

In order to study the thermodynamics and spectroscopy of the system,
we choose a suitable black hole in 2+1 dimensions\cite{carlip}. One
particular class of interesting backgrounds is the charged dilaton
black holes of the Einstein-Maxwell- dilaton theory in which the
dilaton field\cite{gib,pre,gar} is exponentially coupled with the
gauge field, $e^{2 \alpha \phi} F^{2}$.

In the present work, the calculation employs
the proposal of Zeng et al\cite{zeng} considering an outgoing wave which performs periodic
motion outside the horizon. Since the gravity system is periodic with respect to
Euclidean time with a period given by the inverse of the Hawking temperature, it
is assumed that the frequency of the outgoing wave is given by this temperature. Thus the adiabatic invariant quantity can be
found out using the surface gravity term, and using the method of Kunstatter, we can find the area and entropy spectrum.
Recently the area spectrum of BTZ black holes is found by the periodicity method in ref.\cite{alexis} and the spectroscopy of rotating
BTZ black holes is derived in ref.\cite{liu} using the method introduced by Majhi and Vagenas\cite{vagenas}.
 We also investigate the thermodynamics of the charged dilaton black hole. Recently  Akbar et al \cite{akbar} studied
the thermodynamics of charged rotating BTZ black holes.

The  present  paper  is  organized  as  follows. In Sec. 2, we study
the thermodynamics of the charged dilaton black hole in 2+1
dimensions. We calculate the area and entropy spectrum of the system
in Sec. 3. And the results and conclusion of the present study are
summarized in Sec. 4.
\section{Thermodynamics of charged dilaton black hole }
In this work we consider a class of black hole solution obtained
by Chan and Mann\cite{chan}. The solution represents static charged black hole with a
dilaton field. The Einstein-Maxwell-dilaton action considered by Chan and Mann\cite{chan}
 is given below:
\begin{equation}
  S=\int d^3 x\sqrt{-g}[R- \frac{B}{2} (\nabla \phi)^2 + e^{-4a \phi} F_{\mu \nu} F^{\mu \nu}+2 e^{b \phi} \Lambda],
\end{equation}

here $\Lambda$ is treated as the cosmological constant. The constants a, b and B are arbitrary coupling constants. $\phi$ is the dilaton field, $R$ is the
scalar curvature and $F_{\mu \nu}$ is the Maxwell field strength. This action is conformally
related to the low-energy string action in $2+1$ dimensions for $B = 8, b = 4$ and
$a = 1$. These black
holes have very interesting properties.

Exact solutions to this filed equations were found in \cite{chan} and the most general such metric has
two degrees of freedom, and can be written in the form
\begin{equation}\label{gm}
 ds^2=-U(r) dt^2+ \frac{dr^2}{U(r)}+ H^{2}(r) d\theta^2.
\end{equation}
We now consider the ansatz
\begin{equation}
 H^2=\gamma^2 r^{N},
\end{equation}
as a more generic case and further assume that
\begin{equation}
 \phi=k ln(\frac{r}{\beta}),
\end{equation}
\begin{figure}
 \centering
\epsfig{figure=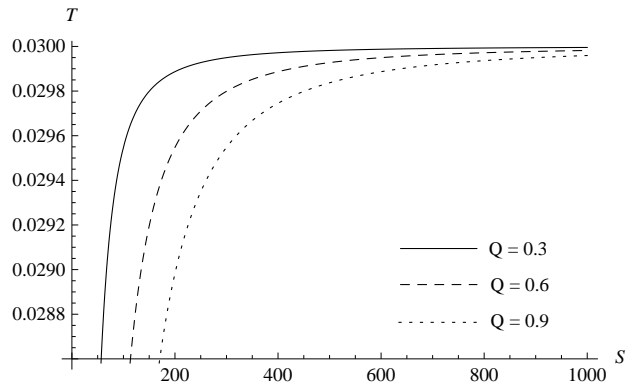}
\caption{Variation of temperature with respect to entropy for different values of charge(Q)}
\label{te}
\end{figure}
where $k$ is a real number.
The solutions of Eq.(\ref{gm}) with the above ansatz depend on the dimensionless couplings a and b ( or alternatively N).
By performing the coordinate transformation $\gamma^2 r^{N}\rightarrow r^2$,
Eq.(\ref{gm}) becomes
\begin{eqnarray}\label{generic}
 ds^2=-\left(A r^{\frac{2}{N}-1}+\frac{8 \Lambda r^2}{(3N-2)N}+\frac{8Q^2}{N(2-N)}\right)dt^2\nonumber\\
 +\frac{4 r^{\frac{4}{N}-1}}{N^2 \gamma^{\frac{4}{N}}\left(A r^{\frac{2}{N}-1}+\frac{8 \Lambda r^2}{(3N-2)N}+\frac{8Q^2}{N(2-N)}\right)}dr^2 +r^2 d
 \theta^2.
\end{eqnarray}
From now on $r$ denotes the usual radial coordinate.
A family of static solutions with rotational symmetry for the action which is mentioned above were thus derived.
With $N=1$, $A=\frac{-2M}{N}=-2M$, Eq.(\ref{generic}) will be
\begin{equation}\label{metric}
 ds^{2}=-(-2Mr+ 8 \Lambda r^{2}+ 8 Q^{2})dt^{2}+ \frac{4 r^{2}dr^{2}}{(-2Mr+ 8 \Lambda r^{2}+ 8 Q^{2})} +r^{2} d\theta^{2}.
\end{equation}
with $\phi= \frac{1}{4} ln(\frac{r}{\beta})$ and $F_{rt}=\frac{Q}{r^{2}}$.
For $M\geq 8 Q\sqrt{\Lambda}$, the space time represents a black hole and its horizons are given by
\begin{figure}
 \centering
\epsfig{figure=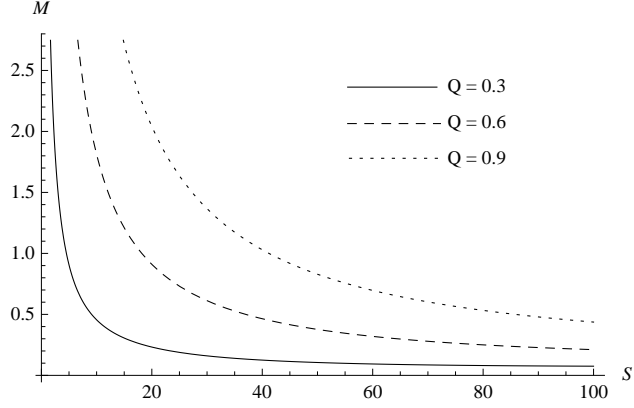}
\caption{Variation of mass of the black hole with respect to the entropy for different values of charge(Q)}
\label{me}
\end{figure}
 \begin{equation}
 r_{+}=\frac{M+\sqrt{M^{2}-64Q^{2} \Lambda}}{8 \Lambda},
\end{equation}
\begin{equation}
 r_{-}=\frac{M-\sqrt{M^{2}-64Q^{2} \Lambda}}{8 \Lambda}.
\end{equation}

We can establish the relation between mass of black hole and horizon radius straightforwardly from the above
equation and is given by
\begin{equation}\label{minr}
 M= \frac{4 Q^{2}}{r_{+}}+ 4 r_{+} \Lambda.
\end{equation}
Redefining this equation in terms of entropy, with the help of the
notion that
 \begin{equation}
        S= 4 \pi r_{+},
 \end{equation}

we  can  find,
\begin{equation}\label{mass}
 M=\frac{S \Lambda}{\pi}+\frac{16 Q^{2}\pi}{S}.
\end{equation}
We can deduce the thermodynamic quantities from the above expression of mass in terms of entropy.
  \begin{equation}\label{temperature}
   T=\left(\frac{\partial M}{\partial S}\right)_{Q},
  \end{equation}
\begin{equation}\label{potential}
 V =\left(\frac{\partial M}{\partial Q}\right)_{S},
\end{equation}
\begin{equation}\label{heatcapacity}
 C_{Q}=T\left(\frac{\partial S}{\partial T}\right)_{Q}.
\end{equation}

\begin{figure}
 \centering
\epsfig{figure=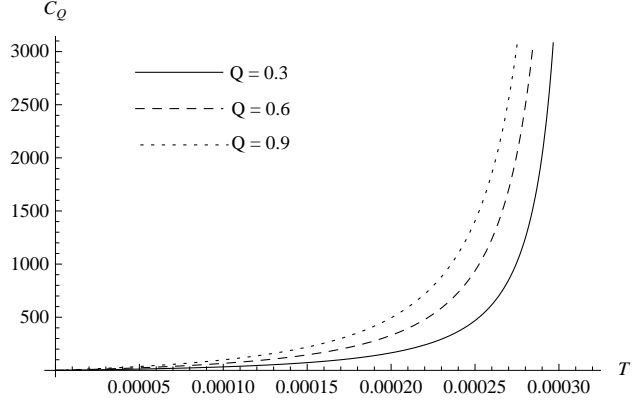}
\caption{Variation of heat capacity with respect to the temperature for different values of charge (Q)}
\label{ht}
\end{figure}

From Eqs. (\ref{mass}) and (\ref{temperature}) we will get the black
hole temperature as,
\begin{equation}
 T=\left(\frac{\partial M}{\partial S}\right)_{Q}=\frac{ \Lambda}{\pi}-\frac{16 Q^{2} \pi}{S^{2}}.
\end{equation}

This is the temperature-entropy relation we obtained and we have plotted the variations in Fig(\ref{te}).
We could see the rapid change of entropy in a certain small interval of temperature. And we take the variation for different charges
and the charge of a back hole is taken in terms of mass.

Fig(\ref{me}) represents the variation of the mass of the charged dilaton black hole with entropy as in Eq.(\ref{mass}).
We take the variation for different values of black hole charge, an
approximate linear relation exists between $M$ and $S$ only asymptotically for $S\gg \frac{Q}{\sqrt{\Lambda}}$.

Now, we look at the potential difference between the horizon and
infinity. That will follow Eq.(\ref{potential}),
\begin{equation}\label{pot}
 V =\left(\frac{\partial M}{\partial Q}\right)_{S}=\frac{32 \pi Q}{S}.
\end{equation}

Replacing the entropy in terms of $r_{+}$,
we obtain
\begin{equation}\label{v}
  V_{+} =\frac{8 Q}{r_{+}}.
\end{equation}

Now the heat capacity  is calculated as in Eq.(\ref{heatcapacity}),
\begin{equation}
 C_{Q}=T\left(\frac{\partial S}{\partial T}\right)_{Q}=\frac{2 Q \pi^{2} T}{(\Lambda -\pi
 T)^{\frac{3}{2}}}.
\end{equation}
The variation of the heat capacity is examined and found that heat
capacity diverges as temperature increases. Heat capacity  of  a
system is positive. According to black hole
thermodynamics\cite{davies}, the more mass and energy a black hole
absorbs, the colder it becomes,  which implies  that Schwarzschild
black hole gets hotter as they radiate out energy since it possesses
a negative heat capacity. Fig(\ref{ht}) represents the variation of
heat capacity with temperature for the charged dilaton black holes.
The heat capacity becomes singular at $T= \frac{\Lambda}{\pi}$. We
have plotted the   graph for different values  of  Q.

\section{Spectroscopy of Charged Dilaton Black hole}

In this part of the work we discuss the properties of entropy
quantization and the area spectrum. We find the area spectrum by
evaluating the adiabatic invariant integral which vary very slowly
compared to variations of the external perturbations of the system.
Given a system with energy E, the first law of  thermodynamics for
charged black hole is\cite{Abbott}:

\begin{equation}\label{1}
dM=\frac{T_{H}}{4}dA+ V dQ.
\end{equation}
We exclusively utilize the period of motion of outgoing wave, which is shown to be related to the vibrational frequency of the
perturbed black hole, to quantize the area of the charged dilaton black hole.
It is well known that the gravity system in Kruskal coordinate is periodic with respect to the Euclidean time. Particle's
motion in this periodic gravity system also owns a period, which has been shown to be the inverse Hawking temperature\cite{Gibbons}.
To  find the area spectrum via the periodicity method, we use Eq.(\ref{metric}) in K G equation,
\begin{equation}\label{kg}
 g^{\mu \nu}\partial_{\mu}\partial_{\nu}\Phi-\frac{m^{2}}{\hbar^{2}}\Phi=0.
\end{equation}
By adopting the wave equation ansatz for the scalar field we can get the solution of wave equation. On the other
hand we can also obtain the solution from the Hamilton-Jacobi equation.
\begin{equation}\label{hj}
  g^{\mu \nu}\partial_{\mu}S\partial_{\nu}S+m^{2}S=0,
\end{equation}
where $S$ is the action and $S$ and $\Phi$ are related by,
\begin{equation}\label{phi}
 \Phi=exp\left[\frac{i}{\hbar}S(t,r)\right].
\end{equation}
The action can be decomposed as\cite{zeng,Chen}
\begin{equation}\label{action}
 S(t,r)= -Et+ W(r),
\end{equation}
where $E$ is the energy of the emitted particle measured by an observer at infinity and (\ref{metric}) gives the function $W$
near the horizon as
\begin{equation}\label{wr}
 W(r)=\frac{i \pi E}{f'(r_{H})},
\end{equation}
where we only consider the outgoing wave near the horizon. In this case, it is obvious that the wave function $\Phi$
outside the horizon can be expressed as the form,
\begin{equation}\label{phipsi}
 \Phi=exp\left[-\frac{i}{\hbar}Et\right] \psi(r_{H}),
\end{equation}
where
\begin{equation}\label{psir}
 \psi(r_{H})=exp\left[-\frac{\pi E}{\hbar f'(r_{H})}\right],
\end{equation}
and from the above equation it is clear that $\Phi$ is a periodic function with period
\begin{equation}\label{tau}
 \tau=\frac{2 \pi \hbar}{E}.
\end{equation}
Using the fact that the gravity system in Kruskal coordinate is
periodic with respect to the Euclidian time, particle's motion in
this periodic gravity system also owns a period, which has been
shown to be the inverse Hawking temperature. Thus the relation
between $\tau$ \& $T$ can be written as
\begin{equation}\label{taut}
 \tau=\frac{2 \pi}{\kappa_{r}}=\frac{\hbar}{T_{H}},
\end{equation}
hence
\begin{equation}\label{tk}
 T_{H}=\frac{\hbar \kappa_{r}}{2 \pi},
\end{equation}
where  $\kappa_{r}$  is  the  surface  gravity  of  the  black hole.
The expression of surface area ( here it will be the circumference
since is is a 2+1 black hole) of the event horizon is
\begin{equation}\label{area}
 A=2\pi r_{+},
\end{equation}
from the above equation the change in the horizon area of a charged dilaton black hole can be written as
\begin{equation}\label{da}
 \Delta A =2 \pi  dr_{+}.
\end{equation}
Now, Eq.(\ref{minr}) gives the differential
\begin{equation}\label{dm}
 dM=\left[4 \Lambda- \frac{4 Q^{2}}{r_{+}^{2}}\right]dr_{+} + \frac{8Q}{r_{+}} dQ.
\end{equation}
Using the lapse function we can calculate the surface gravity as
follows
\begin{equation}\label{kappa}
 \kappa_{r_{+}}=\frac{1}{2} \frac{df(r)}{dr}|_{r=r_{+}}=\left[2 \Lambda-\frac{2
 Q^{2}}{r_{+}^{2}}\right],
\end{equation}
where $ f(r)=-M+\frac{4Q^{2}}{r_{+}}+ 4 \Lambda r_{+}$. Using
Eq.(\ref{kappa}) in Eq.(\ref{dm})
\begin{equation}\label{newdm}
 dM= 2 \kappa_{r_{+}}dr_{+} + \frac{8Q}{r_{+}} dQ.
\end{equation}
 and substituting Eq.(\ref{v})
\begin{equation}
 dr_{+}= \frac{dM - V_{+}dQ}{2 \kappa_{r_{+}}}.
\end{equation}
Now we can use Eq.(\ref{da})
\begin{equation}
 \Delta A = \pi \left[\frac{dM - V_{+}dQ}{ \kappa_{r_{+}}}\right].
\end{equation}

Now, the adiabatic invariant integral for the charged 2+1 black hole
can be written as
\begin{equation}
 I=\int\pi\frac{dM - V_{+} dQ}{\kappa_{r_{+}}}.
\end{equation}
Now we can write the surface gravity, $\kappa_{r_{+}}$ in terms of $M$, $Q$ and  $r_{+}$ to carry out the integration,
and is given by
\begin{equation}
 \kappa_{r_{+}}=\frac{\sqrt{M^{2}-64 Q^{2} \Lambda}}{2 r_{+}}.
\end{equation}
 The adiabatic invariant integral will read as
\begin{equation}
 I=\int\frac{dM - V_{+} dQ}{\frac{\sqrt{M^{2}-64 Q^{2} \Lambda}}{2 r_{+}}}.
\end{equation}

The integral can be solved and using the value of the potential
given in Eq.(\ref{pot})
\begin{equation}
 I=\frac{1}{4\Lambda} \int \frac{M}{\sqrt{M^{2}-64 Q^{2} \Lambda}}dM + \frac{1}{4 \Lambda} \int dM - 16 \int \frac{Q}{\sqrt{M^{2}-64 Q^{2} \Lambda}}dQ.
\end{equation}
Integrating, we get
\begin{equation}
 I=\frac{M+2\sqrt{M^{2}-64 Q^{2}}}{4 \Lambda}.
\end{equation}
Replacing $M$ in terms of $r_{+}$ using Eq.(\ref{minr}) and applying Bohr-Sommerfeld quantization condition we find
\begin{equation}
 I= 3 r_{+} -\frac{Q^{2}}{\Lambda r_{+}}\simeq n \hbar.
\end{equation}
 Finding the circumference term, $2 \pi r_{+}$(since this is a 2+1 black hole), in the above expression
we can write the circumference spectrum as

\begin{equation}
 A_{n}  \simeq \frac{2 \pi n \hbar}{3} -\frac{2 \pi Q^{2}}{3 \Lambda r_{+}}.
\end{equation}

Thus we see that the circumference of the charged dilaton black hole
is discrete and the spacing is equidistant. In  this system, the
circumference spectrum depends on the black hole parameters,  but
the 'circumference spacing' is equal and is independent of the black
hole parameters. Recalling that the black hole entropy is
proportional to the black hole horizon area(here it is the
circumference), it is clear that the entropy is also quantized.

\section{SUMMARY AND CONCLUSION}
We have investigated the thermodynamic and spectroscopic aspects of
charged dilaton black hole in 2+1 dimensions. We have obtained the
variation of entropy with temperature, mass with entropy and heat
capacity with temperature. The heat capacity is found to be
positive. We have found the circumference spectrum. It is found that
the though the circumference spectrum is black hole parameter
dependent, the circumference spacing is independent of the black
hole parameters.
\section*{Acknowledgments}
The  authors  are  thankful to  the  Reviewers  for  valuable
suggestions. TR wishes to thank UGC, New Delhi for financial support
under RFSMS scheme. VCK is thankful to UGC, New Delhi for financial
support through a Major Research Project and wishes to acknowledge
Associateship of IUCAA, Pune, India.



\begin{thebibliography}{0}
\bibitem{Bekenstein} J. D. Bekenstein, Lett . Nuovo Cimento Soc. Ital. fis. \textbf{11} 467 (1974)
\bibitem{Hod} S. Hod Phys, Rev. Lett \textbf{81} 4293(1998).
\bibitem{Kunstatter} G.Kunstatter, Phys.Rev.Lett. \textbf{90} 161301(2003).
\bibitem{mukhanov} J. D. Bekenstein and V. F. Mukhanov, Phys. Lett. B \textbf{360}, 7 (1995)
\bibitem{chan}K. C. K. Chan, R. B. Mann , Phys. Rev. D50 (1994) 6385; erratum, D52 2600 (1995)
\bibitem{Maggiore}M . Maggiore, Phys. Rev. Lett. \textbf{100}, 141301 (2008)
\bibitem{Songhai} Songhai Chen et al , IOP , Class Quan Grav \textbf{22}(2005)
\bibitem{carlip} S. Carlip, Class. Quantum Grav. \textbf{12} (1995) 2853–2879.
\bibitem{zeng} X. Zeng, X. Liu and W. Liu, Eur. Phys. J. C \textbf{72}(2012).
\bibitem{alexis} Alexis Larranaga, Int. J. Mod. Phys. D, \textbf{21}, 1250068 (2012).
\bibitem{liu} X. Liu, X. Zeng and W. Liu, Sci China-Phys Mech Astron, \textbf{55}, 1747- 1750 (2012).
\bibitem{vagenas} B.R. Majhi, E.C. Vagenas, Phys. Lett. B \textbf{701}, 623 (2011).
\bibitem{akbar} M. Akbar et al, Phys. Rev. D \textbf{83}, 084031 (2011).
\bibitem{gib} G. W. Gibbons et al Nucl. Phys. B \textbf{298}  741 (1988).
\bibitem{pre} J. Preskill et al Mod. Phys. Lett. A \textbf{6}  2353 (1991).
\bibitem{gar} D. Garfinkle et al Phys. Rev. D \textbf{43}  3140 (1991).
\bibitem{davies} P. C. W. Davies, Proc. R. Soc. Loud. A. \textbf{353}, 499-521 (1977).
\bibitem{Chen} D. Y. Chen et al, Phys. Lett. B 665(2008) 106.
\bibitem{Gibbons} G. W. Gibbons, M.J Perry, Proc. R. Soc. Lond A \textbf{358}, 467(1978).
\bibitem{Abbott} L.F.Abbott et al., Nucl. Phys. B \textbf{195}  76(1982).


\end{thebibliography}
\end{document}